\documentclass[journal]{IEEEtran}
\usepackage{times}
\usepackage{soul}
\usepackage{url}
\usepackage[hidelinks]{hyperref}
\usepackage[utf8]{inputenc}
\usepackage[small]{caption}
\usepackage{graphicx}
\usepackage{amsmath}
\usepackage{amsthm}
\usepackage{booktabs}
\usepackage[switch]{lineno}

\usepackage{multirow}
\usepackage{amsfonts}
\usepackage{colortbl}
\usepackage{xcolor}
\usepackage{makecell}
\usepackage{comment}
\usepackage{eqparbox}
\usepackage{subfigure}
\usepackage{color}
\usepackage{array}
\usepackage{bm}
\usepackage[symbol]{footmisc}
\usepackage[nolist,nohyperlinks]{acronym}
\usepackage[linesnumbered,ruled,vlined]{algorithm2e}

\acrodef{FL}{Federated Learning}
\acrodef{ARIMA}{Autoregressive Integrated Moving Average}
\acrodef{ML}{Machine Learning}
\acrodef{SVM}{Support Vector Machine}
\acrodef{DT}{Decision Tree}
\acrodef{RF}{Random Forest}
\acrodef{DL}{Deep Learning}
\acrodef{RNN}{Recurrent Neural Network}
\acrodef{LSTM}{Long Short-Term Memory}
\acrodef{GRU}{Gated Recurrent Unit}
\acrodef{CNN}{Convolutional Neural Network}
\acrodef{GAN}{Generative Adversarial Network}
\acrodef{MLP}{Multi-Layered Perceptron}
\acrodef{FFT}{Fast Fourier Transform}
\acrodef{CSTI}{Cross-Stock Trend Integration}
\acrodef{DFT}{Discrete Fourier Transform}
\acrodef{MSE}{Mean Squared Error}
\acrodef{NLP}{Natural Language Processing}
\acrodef{MHSA}{Multi-Head Self-Attention}
\acrodef{FFN}{Feed-Forward Networks}
\acrodef{FNSPID}{Financial News and Stock Price Integration Dataset}
\acrodef{SGD}{Stochastic Gradient Descent}

\begin{document}

\title{From Local Patterns to Global Understanding: Cross-Stock Trend Integration for Enhanced Predictive Modeling}

\author{Yi~Hu,~
        Hanchi~Ren*,~
        Jingjing~Deng,~
        and~Xianghua~Xie,~\IEEEmembership{IEEE Senior Member}
\thanks{Y. Hu, H. Ren and X. Xie are with the Department of Computer Science, Swansea University, United Kingdom.\\

J. Deng is with the Department of Computer Science, Durham University, United Kingdom.\\

Email: hanchi.ren@swansea.ac.uk}}


\maketitle

\begin{abstract}
    Stock price prediction is a critical area of financial forecasting, traditionally approached by training models using the historical price data of individual stocks. While these models effectively capture single-stock patterns, they fail to leverage potential correlations among stock trends, which could improve predictive performance. Current single-stock learning methods are thus limited in their ability to provide a broader understanding of price dynamics across multiple stocks. To address this, we propose a novel method that merges local patterns into a global understanding through cross-stock pattern integration. Our strategy is inspired by Federated Learning (FL), a paradigm designed for decentralized model training. FL enables collaborative learning across distributed datasets without sharing raw data, facilitating the aggregation of global insights while preserving data privacy. In our adaptation, we train models on individual stock data and iteratively merge them to create a unified global model. This global model is subsequently fine-tuned on specific stock data to retain local relevance. The proposed strategy enables parallel training of individual stock models, facilitating efficient utilization of computational resources and reducing overall training time. We conducted extensive experiments to evaluate the proposed method, demonstrating that it outperforms benchmark models and enhances the predictive capabilities of state-of-the-art approaches. Our results highlight the efficacy of Cross-Stock Trend Integration (CSTI) in advancing stock price prediction, offering a robust alternative to traditional single-stock learning methodologies.
\end{abstract}


\IEEEpeerreviewmaketitle

\section{Introduction}

\IEEEPARstart{S}{tock} price prediction has long been a cornerstone of financial research, with its origins rooted in attempts to model and forecast market behavior for informed decision-making. Early efforts in stock price prediction were dominated by statistical models, such as \ac{ARIMA}~\cite{box1970distribution,lee2011forecasting,funde2023comparison} and exponential smoothing~\cite{gardner1985exponential,gardner2006exponential,de2009predicting} methods, which relied on historical price data to identify patterns and trends. These models were effective within their time but were limited by their linear assumptions and inability to capture the complexities of financial markets. 

\begin{figure}[t!]
    \centering
    \includegraphics[width=0.99\linewidth]{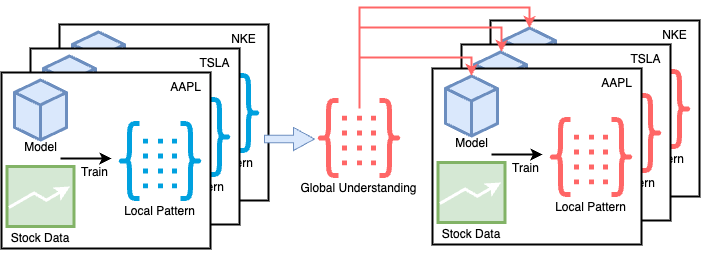}
    \caption{Illustration of CSTI}
    \label{fig:overview}
\end{figure}

The advent of computational advancements and the rise of \ac{ML} brought a paradigm shift to stock price prediction. Techniques such as \ac{SVM}~\cite{hearst1998support}, \ac{DT}~\cite{de2013decision}, and ensemble methods like \ac{RF}~\cite{breiman2001random} gained popularity for their ability to model nonlinear relationships in financial data. With the availability of large datasets and increased computational power, \ac{DL} approaches, including \ac{RNN}~\cite{elman1990finding}, \ac{LSTM}~\cite{hochreiter1997long}, \ac{GRU}~\cite{cho2014learning}, and Transformer~\cite{vaswani2017attention}, emerged as dominant methods. These models excelled at capturing temporal dependencies and intricate patterns, significantly improving prediction accuracy. Despite these advancements, most traditional \ac{ML} and \ac{DL} approaches focus on training models using individual stock data, often neglecting the inter-dependencies between stocks. Based on these traditional approaches, many novel methods are proposed aiming to achieve higher accuracy of prediction, such as, FilterNet~\cite{yi2024filternet}, FreTS~\cite{yi2024frequency}, DLinear~\cite{zeng2023transformers}, TimesNet~\cite{wu2022timesnet} and PatchTST~\cite{nie2022time}. FilterNet is tailored for time series forecasting, emphasizing the extraction of informative temporal patterns through learnable frequency filters. By selectively amplifying or attenuating specific components of time series signals, FilterNet effectively captures both high-frequency and low-frequency information. This design addresses challenges such as vulnerability to high-frequency noise and inefficiencies in full-spectrum utilization, which are common in Transformer-based models. FreTS introduces a novel approach by applying \ac{MLP} in the frequency domain. This method involves transforming time-domain signals into their frequency-domain representations using the \ac{DFT}. FreTS then employs redesigned \ac{MLP} to learn the real and imaginary parts of these frequency components, capturing global dependencies and concentrating on key frequency components with compact signal energy. Operating on both inter-series and intra-series scales, FreTS effectively learns channel-wise and time-wise dependencies. DLinear is a simple and fast model for long-horizon forecasting that utilizes linear layers for trend and seasonality components, offering competitive accuracy with reduced complexity. TimesNet is a \ac{CNN}-based model that transforms time series data into 2D tensors using the \ac{FFT}, enabling the application of visual backbones like inception to capture temporal patterns effectively. PatchTST is a Transformer-based model that employs a patch-based technique inspired by computer vision, achieving state-of-the-art prediction results in long-term time series forecasting. While those approaches have produced good results, this single-stock perspective can cause the model to miss broader market trends or cross-stock relationships that can provide valuable predictive insights. Researchers have begun exploring collaborative and integrative strategies to address these limitations~\cite{yan2023cross}. For example, K. J. Koa et al.~\cite{koa2023diffusion} introduced a model that combines deep hierarchical variational auto-encoders with diffusion probabilistic techniques to predict stock prices over multiple steps, addressing the inherent stochastic nature of stock data; Z. Pei et al.~\cite{pei2024stock} proposed a method that leverages time series decomposition and multi-scale convolutional neural networks to predict stock prices using Open, High, Low, Close, and Trading volume data from multiple stocks; Y. Dong and Y. Hao~\cite{dong2024stock} presented a deep neural network framework that dynamically assigns weights to multidimensional features from various stocks, capturing the impact of each feature on stock prices; W. Liu et al.~\cite{liu2024multi} introduced a method that calculates multiple factors, including Alpha158 and OCHLVC (Open, Close, High, Low, Volume, and Change) data, to predict stock prices using a \ac{GAN}~\cite{goodfellow2020generative} combined with TrellisNet~\cite{bai2018trellis}.

In recent years, the fusion of financial domain knowledge with sophisticated computational techniques has opened new avenues for stock price prediction. The incorporation of external data sources, such as news sentiment~\cite{dong2024fnspid}, macroeconomic indicators~\cite{lee2023effective}, and social media trends~\cite{nyakurukwa2023evolution}, alongside advancements in model architectures, continues to push the boundaries of what is possible in financial forecasting. As markets grow increasingly complex and interconnected, methods that transcend traditional single-stock modeling and embrace holistic, cross-stock approaches represent a promising direction for future research. The essence of \ac{FL} lies in merging local models to create a global model that captures the latent feature space of each local dataset without compromising the privacy or exposing the details of any individual dataset. This concept inspired our approach, as we recognized that the trends of different stocks could be similarly merged to form a unified global understanding. Stock price movements, despite their unique characteristics, often share underlying latent patterns. By leveraging this idea, we aim to enhance prediction performance by integrating these shared patterns into a cohesive global model. In this work, we propose a novel training strategy designed to transform local patterns derived from individual stocks into a cohesive global understanding, thereby enhancing predictive modeling through the \ac{CSTI}. Specifically, our approach involves training individual models on the historical price data of different stocks and then merging these models iteratively during training. This iterative merging process allows the global model to progressively capture shared patterns and relationships among stocks, fostering a comprehensive understanding of cross-stock price dynamics. Once the global model achieves a sufficient level of convergence, we fine-tune it on the data of each individual stock to adapt and optimize its predictions for specific stocks. As shown in Figure~\ref{fig:overview}, this two-phase training strategy not only preserves the unique characteristics of each stock but also leverages the broader market relationships that enhance prediction accuracy. By integrating cross-stock trends into the learning process, our method addresses the limitations of traditional single-stock training and offers a robust framework for improving stock price prediction models. Our strategy inherently supports parallel training, allowing individual stock models to be trained simultaneously, which significantly improves computational efficiency and scalability. We have conducted comprehensive experiments on famous benchmark dataset, \ac{FNSPID}, and evaluated our approach using multiple state-of-the-art models to validate its effectiveness. The results illustrate that our proposed training strategy significantly enhances overall performance, outperforming traditional single-stock training methods and existing benchmark models. By integrating cross-stock trends, our method not only improves prediction accuracy but also demonstrates robustness across different market conditions and datasets.

\section{Related Work}
\label{sec:rw}

\subsection{Federated Learning}
Personal data protection and privacy-preserving issues have garnered significant attention from researchers~\cite{liu2020survey,tanuwidjaja2020privacy,koti2021swift}. Traditional machine learning approaches, which typically require centralized data for model training, are becoming increasingly challenging to implement due to stringent restrictions on data sharing. As a result, decentralized data-training approaches have emerged as a more attractive alternative, offering notable advantages in privacy preservation and data security. \ac{FL}~\cite{konevcny2016federated,mcmahan2017communication} was introduced as a solution to these concerns, enabling individual data providers to collaboratively train a shared global model without the need for centralized data aggregation. ~\cite{mcmahan2017communication} proposed a practical decentralized training method for deep neural networks based on averaging aggregation. Their experimental evaluations, conducted on a variety of datasets and architectures, demonstrated the robustness and effectiveness of \ac{FL} in addressing privacy and security challenges.

\subsection{Stock Trend Prediction}

Traditionally, statistical approaches have been developed for time series forecasting, focusing on both the time and frequency domains. In recent years, \ac{DL} methods have gained popularity in this field, owing to their ability to capture nonlinear and complex relationships. These techniques leverage architectures such as \ac{RNN}, \ac{LSTM}, \ac{GRU} and \ac{CNN} to model dependencies within either the time or frequency domain. Furthermore, Transformer-based forecasting models have emerged as a powerful alternative, utilizing attention mechanisms to effectively model long-range dependencies.

\textbf{FilterNet} is designed to enhance time series forecasting by leveraging learnable frequency filters. It can be expressed as:
\begin{align}
    \mathcal{Z} &= \mathcal{F}^{-1}(\mathcal{F}(Z) \cdot \mathcal{H}_{filter}) \nonumber
\end{align}
where $\mathcal{F}$ is Fourier Transform, $\mathcal{F}^{-1}$ is inverse Fourier Transform and $\mathcal{H}_{filter}$ is learnable frequency filter. It addresses challenges such as vulnerability to high-frequency signals and inefficiencies in full-spectrum utilization, which are common in Transformer-based models. FilterNet introduces two types of learnable filters: plain shaping filter (PaiFilter) and contextual shaping filter (TexFilter). The first one adopts a universal frequency kernel $\mathcal{H}_{Pai}$ for signal filtering and temporal modeling. $\mathcal{H}_{Pai}$ is a random initialized learnable weight. The PaiFilter processes the input time series to selectively pass or attenuate certain frequency components, effectively capturing essential temporal patterns.
\begin{align}
    \mathcal{Z} &= \mathcal{F}^{-1}(\mathcal{F}(Z) \cdot \mathcal{H}_{Pai}) \nonumber
\end{align}
The second examines filtered frequencies in terms of their compatibility with input signals for dependency learning. It adapts to the specific context of the input data, allowing for more nuanced modeling of temporal dependencies. 
\begin{align}
    \mathcal{Z} &= \mathcal{F}^{-1}(\mathcal{F}(Z) \cdot \mathcal{H}_{Tex}(\mathcal{F}(Z))) \nonumber
\end{align}
where $\mathcal{H}_{Tex}$ is a neural network acts as data-dependent frequency filter. FilterNet can approximate the linear and attention mappings widely adopted in time series literature. It effectively handles high-frequency noises and utilizes the entire frequency spectrum beneficial for forecasting. Extensive experiments on eight time series forecasting benchmarks have demonstrated FilterNet’s superior performance in both effectiveness and efficiency compared with state-of-the-art methods.

\textbf{FreTS} applies \ac{MLP} in the frequency domain and involves transforming time-domain signals into their frequency-domain representations using the \ac{DFT}:
\begin{align}
    \mathcal{X}(f) &= \sum^{T-1}_{t=0}x(t) \cdot e^{-j2\pi ft} \nonumber
\end{align}
where $x(t)$ is the time-series data, $e^{-2j\pi ft}$ is an exponential function that decomposes the signal into its frequency components. FreTS then employs redesigned \ac{MLP} to learn the real ($Re(\mathcal{X})$) and imaginary ($Im(\mathcal{X})$) parts of these frequency components, capturing global dependencies and focusing on key frequency components with compact signal energy:
\begin{align}
    \hat{\mathcal{X}}(f) &= MLP(Re(\mathcal{X})) + j \cdot MLP(Im(\mathcal{X})) \nonumber
\end{align}
Operating on both inter-series and intra-series scales, FreTS effectively learns channel-wise and time-wise dependencies. Experiments on 13 real-world benchmarks, encompassing both short-term and long-term forecasting tasks, have shown that FreTS consistently outperforms many existing methods.

\textbf{DLinear} is a time series forecasting model designed to address the challenges of modeling both long-term trends and short-term seasonality in temporal data. It emphasizes the importance of decomposing time series data into two components: trend and seasonal, and then applying linear modeling to each component separately for enhanced forecasting accuracy. The decomposition can be expressed as:
\begin{align}
    x(t) = x_{trend}(t) + x_{seansonal}(t) \nonumber
\end{align}
where $x(t)$ is the original time series, $x_{trend}(t)$ captures the long-term trends, and $x_{seansonal}(t)$ represents the short-term periodic fluctuations. For the trend component, a simple linear model is used to approximate the general upward or downward trajectory over time. The linear trend is modeled as:
\begin{align}
    x_{trend}(t) = w_{trend} \cdot t + b_{trend} \nonumber
\end{align}
where $w_{trend}$ is the weight (slope) capturing the direction and magnitude of the trend, and $b_{trend}$ is the bias term. For the seasonal component, periodic patterns are captured using a Fourier decomposition approach, which breaks the time series into sinusoidal components. The seasonal component can be modeled as:
\begin{align}
    x_{seansonal}(t) = \sum_{k = 1}^{K}(a_{k}\cdot cos(\frac{2\pi kt}{T}) + b_{k} \cdot sin(\frac{2\pi kt}{T})) \nonumber
\end{align}
where $K$ is the number of harmonics (frequency components), $T$ is the period of the seasonality (e.g., daily, weekly, etc.), and $a_{k}, b_{k}$ are learnable coefficients representing the amplitude of the cosine and sine waves, respectively. This decomposition allows the model to adapt to various periodic patterns in the data. One of DLinear’s key strengths is its simplicity and focus on linear relationships, which makes it computationally efficient compared to more complex deep learning models. By separating the linear components explicitly, DLinear avoids the overfitting risks often associated with large neural networks while retaining the ability to capture essential temporal patterns. Additionally, the model can be trained in parallel for each component, enhancing scalability.

\textbf{TimesNet} is an advanced model for time series forecasting that innovatively transforms 1D time series data into 2D tensors, enabling it to capture both intra-period and inter-period variations effectively. This approach is particularly beneficial for modeling complex temporal patterns, including periodicity and irregular trends. The transformation involves reshaping the time series data $x(t)$ into a 2D tensor $X$ based on a given period $P$, defined as:
\begin{align}
    X[i,j] = x(t+i\cdot P + j) \nonumber
\end{align}
where $i$ indexes the inter-period variations (across different cycles of the period $P$), and $j$ captures intra-period variations (within a single cycle). This reshaping allows the model to treat periodic dependencies as spatial patterns, which can be efficiently processed using 2D convolution operations. The core building block of TimesNet is the TimesBlock, which employs parameter-efficient 2D convolutions to model temporal variations. For a given tensor input $X$ , the TimesBlock applies a series of 2D convolution layers:
\begin{align}
    F = \sigma(Conv2D(X,W) + b) \nonumber
\end{align}
where $W$ and $b$ are the learnable weights and biases of the convolution kernel, and $\sigma$ represents an activation function such as ReLU. The convolution operation captures local temporal dependencies by aggregating information across neighboring intra-period and inter-period variations. To adaptively discover multi-periodicity in time series data, TimesNet incorporates an inception-like architecture within the TimesBlock, enabling it to process multiple kernel sizes simultaneously. This allows the model to extract features at different temporal resolutions. For instance, convolutions with larger kernels capture long-term dependencies, while smaller kernels focus on short-term variations. Additionally, TimesNet includes a periodicity discovery mechanism to identify the most relevant period $P$ for each dataset. This mechanism involves analyzing the autocorrelation of the time series and selecting the period with the highest periodicity score, ensuring that the reshaped tensor optimally represents temporal variations. The output of TimesNet is aggregated using global pooling and fed into a fully connected layer for prediction. The model is trained using a loss function, such as \ac{MSE}.

\textbf{PatchTST} is a Transformer-based architecture specifically designed for time series forecasting. It introduces the concept of dividing time series data into non-overlapping patches, treating these patches as analogous to words in \ac{NLP} tasks. This patching mechanism helps capture both local and global temporal dependencies efficiently. Let $x \in \mathbb{R}^{T \times d}$ represent a multivariate time series with $T$ time steps and $d$ features. PatchTST segments $x$ into $N = \frac{T}{p}$ patches of size $p$, such that each patch $P_i$ is defined as $P_i = \{x_{t_i}, x_{t_{i+1}}, \ldots, x_{t_{i+p-1}}\}$. These patches are then linearly embedded into a feature space using a learnable embedding matrix $W_e$:
\begin{align}
Z_i = P_i \cdot W_e + b_e \nonumber
\end{align}
where $W_e \in \mathbb{R}^{p \times d_z}$ maps each patch into a $d_z$-dimensional feature space, and $b_e$ is a bias term. The patches are then passed through a Transformer encoder, where the self-attention mechanism plays a pivotal role. The attention mechanism operates as:
\begin{align}
\text{Attention}(Q, K, V) = \text{softmax}\left(\frac{QK^\top}{\sqrt{d_k}}\right)V \nonumber
\end{align}
where $Q = ZW_q$, $K = ZW_k$, and $V = ZW_v$. Here, $W_q, W_k, W_v \in \mathbb{R}^{d_z \times d_k}$ are learnable projection matrices for queries, keys, and values, respectively, and $d_k$ is the dimensionality of the key vectors. The scaled dot-product attention computes the relationship between all patches, allowing the model to capture both short-term and long-term dependencies across the time series. To enhance the model’s ability to focus on temporal patterns, positional encodings $E$ are added to the patch embeddings $Z$, ensuring that the Transformer is aware of the order of patches:
$\tilde{Z} = Z + E$. The Transformer encoder processes these embeddings through multiple layers of \ac{MHSA} and \ac{FFN}. Each encoder block can be expressed as:
\begin{align}
    \tilde{Z}^{\prime} &= \text{LayerNorm}(\tilde{Z} + \text{MHSA}(\tilde{Z})) \nonumber\\
    Z^{\prime \prime} &= \text{LayerNorm}(\tilde{Z}^{\prime} + \text{FFN}(\tilde{Z}^{\prime})) \nonumber
\end{align}
The final output $Z^{\prime \prime}$ is flattened and passed through a regression head for forecasting. PatchTST’s design, particularly the use of patching and Transformers, enables it to model both local temporal structures within patches and global dependencies across patches effectively. 

While the aforementioned methods address various aspects of time series forecasting, including frequency-domain analysis (FilterNet, FreTS), linear modeling (DLinear), 2D variation modeling (TimesNet), and patch-based Transformer architectures (PatchTST), our approach sets itself apart by focusing on the integration of cross-stock patterns to enhance stock price prediction. Drawing inspiration from \ac{FL} principles, our method merges local models trained on individual stocks into a unified global model, effectively capturing inter-dependencies among different stocks without the need for centralized data aggregation. This approach enables the model to uncover shared patterns across stocks, potentially improving predictive performance in stock price forecasting. Moreover, our proposed strategy is highly versatile and can be seamlessly deployed on top of all the aforementioned models. By doing so, our method complements and enhances existing approaches, enabling them to learn cross-stock patterns and overcome their current limitations. This synergy has the potential to unlock even greater performance in stock price prediction tasks.

\begin{figure*}[ht!]
    \centering
    \includegraphics[width=0.99\linewidth]{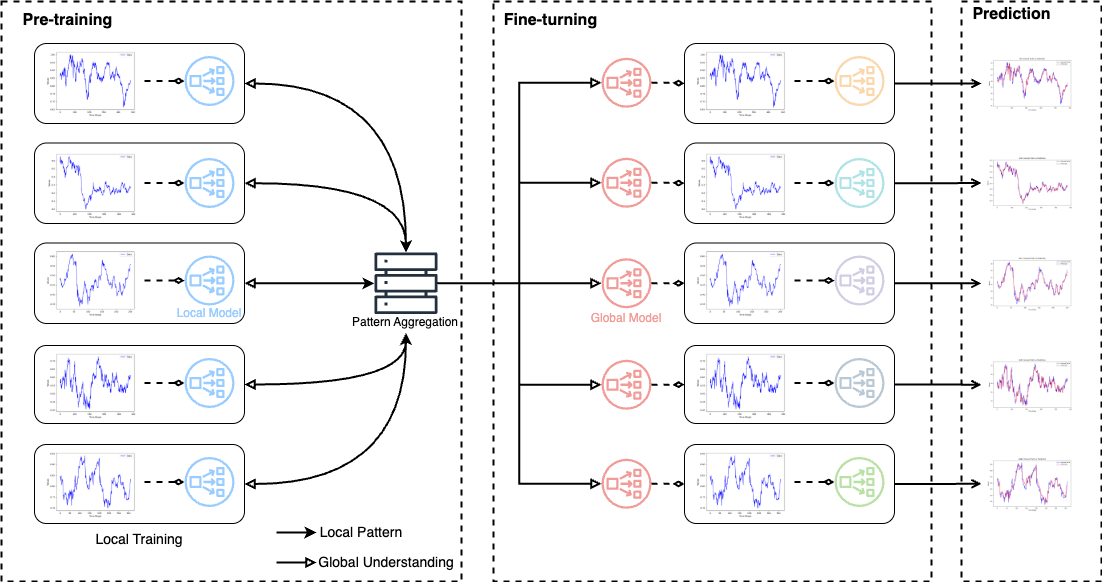}
    \caption{The overview architecture of our proposed CSTI.}
    \label{fig:archi}
\end{figure*}

\section{Proposed Method}
\label{sec:pm}

In this work, we propose a novel training strategy, \ac{CSTI}, which enhances stock price prediction by leveraging shared patterns and interdependencies among stocks. As shown in Figure~\ref{fig:archi}, unlike traditional single-stock forecasting methods that rely solely on individual stock data, \ac{CSTI} extracts local patterns from each stock and integrates them into a unified global understanding, enabling a more holistic representation of market dynamics. The proposed framework consists of three key phases: individual training, global model merging, and fine-tuning, each contributing to the development of a robust, generalizable model for stock prediction. A pseudocode is provided in Algorithm~\ref{alg:csti}.

\begin{algorithm}[ht!]
\caption{CSTI Framework}
\label{alg:csti}
\KwIn{Stock datasets $\{\mathbf{X}_k, \mathbf{y}_k\}_{k=1}^K$, number of stocks $K$, training epochs $E$, global learning rate $\eta_g$, adaptation coefficient $\alpha$, regularization weight $\lambda$}

\KwOut{Fine-tuned models $\{f_k(\cdot; \theta_k')\}_{k=1}^K$ for each stock}

\textbf{Phase 1: Individual Training}

\ForAll{stock $S_k$, where $k = 1$ to $K$ \textbf{in parallel}}{

    Initialize model $f_k(\cdot; \theta_k)$
    
    
    Update $\theta_k$ by minimizing:
        
    $\mathcal{L}_k = \frac{1}{T} \sum_{t=1}^{T} (f_k(\mathbf{X}_{k,t}; \theta_k) - y_{k,t})^2$
}

\textbf{Phase 2: Global Model Merging}

\ForAll{stock $S_k$}{

    Compute weight $w_k$
    
}

Aggregate global model: $\theta_g = \frac{1}{K} \sum_{k=1}^{K} w_k \theta_k$

\For{$e = 1$ to $E$}{

    Refine $\theta_g$ by repeating Phase 1 \& 2.
    
}

\textbf{Phase 3: Fine-Tuning for Stock-Specific Adaptation}

\ForAll{stock $S_k$ \textbf{in parallel}}{

    Initialize $\theta_k \gets \theta_g$
    
    
    \For{$e = 1$ to $E$}{
    
        Update $\theta_k$ by minimizing:
        
        $\mathcal{L}_k' = \frac{1}{T} \sum_{t=1}^{T} (f_k(\mathbf{X}_{k,t}; \theta_k) - y_{k,t})^2 + \lambda \|\theta_k - \theta_g\|^2$
        
    }
    
    Save $\theta_k'$
}

\Return Fine-tuned models $\{f_k(\cdot; \theta_k')\}_{k=1}^K$
\end{algorithm}

\subsection{Individual Training} 
The first phase of the \ac{CSTI} framework focuses on training individual models for each stock using historical stock price data. Each stock, denoted as $ S_k $, is represented by its time-series dataset $ \mathbf{X}_k \in \mathbb{R}^{T \times d} $, where $ T $ represents the length of the time horizon and $ d $ denotes the number of extracted features for each time step. The goal of this phase is to enable each stock-specific model to learn its own price patterns, volatilities, and dependencies on various market indicators. To train a local model $ f_k(\cdot; \theta_k) $ for stock $ S_k $, we define an objective function that optimizes the model parameters $ \theta_k $ to minimize the discrepancy between the predicted stock prices and their actual observed values. The training process is based on the following loss function:
\begin{align}
\mathcal{L}_k = \frac{1}{T} \sum_{t=1}^{T} \ell(f_k(\mathbf{X}_{k, t}; \theta_k), y_{k, t}),
\end{align}
where $ \ell $ represents the stock price prediction loss function, $ \mathbf{X}_{k, t} $ denotes the feature set at time $ t $, and $ y_{k, t} $ represents the actual stock price at time $ t $. The loss function is typically designed as the \ac{MSE} to penalize large deviations in predictions:
\begin{align}
\ell(f_k(\mathbf{X}_{k, t}; \theta_k), y_{k, t}) = (f_k(\mathbf{X}_{k, t}; \theta_k) - y_{k, t})^2.
\end{align}
To ensure robustness, each local model is trained using a combination of historical market indicators, technical indicators, and external financial variables. The feature set $ \mathbf{X}_{k, t} $ may include open price, close price, trading volume, moving averages, volatility measures, and macroeconomic indicators. Feature engineering techniques, such as time-series normalization, log transformation, and seasonal adjustment, are applied to preprocess the data, ensuring stability in learning patterns. A key advantage of this phase is that training is conducted in parallel for all stocks, allowing for efficient scalability. Given a set of $ K $ stocks, their respective models $ \{f_k\}_{k=1}^{K} $ are trained simultaneously on independent computing units, significantly reducing the computational overhead:
\begin{align}
\mathcal{L} = \sum_{k=1}^{K} \mathcal{L}_k.
\end{align}
Parallel training enables the framework to handle a vast number of stocks without compromising computational efficiency. It also allows for the incorporation of multiple asset types, including equities, commodities, and exchange-traded funds, making the system flexible for large-scale financial modeling. Each local model learns stock-specific dependencies, capturing intrinsic price movement trends, cyclical patterns, and market anomalies. By training independently, the models are optimized to their respective stock behaviors, avoiding cross-stock contamination during initial training. This ensures that local training preserves fine-grained stock-specific insights before transitioning to the global model merging phase, where broader inter-stock relationships are integrated.

\subsection{Global Model Merging} 
In the second phase, the locally trained models are iteratively merged to create a global model $ f_g(\cdot; \theta_g)$. This merging process integrates the parameter spaces $ \{\theta_k\}_{k=1}^{K}$ of individual stock models, forming a generalized representation of cross-stock trends. Unlike single-stock models that learn in isolation, this global model captures inter-stock dependencies, enabling a more robust and holistic understanding of market-wide movements.  To achieve this, the global model parameters $ \theta_g$ are computed using a weighted aggregation of the locally trained models:
\begin{align}
\theta_g = \frac{1}{K} \sum_{k=1}^{K} w_k \theta_k,
\end{align}
where $ w_k$ is a weighting factor assigned to stock $ S_k$ based on predefined metrics such as stock volatility, market capitalization, or historical prediction accuracy. This weighted aggregation ensures that more reliable and stable stocks contribute more significantly to the global model, while highly volatile or less predictive stocks have a reduced impact. The adaptive weighting mechanism prevents dominant stocks from disproportionately influencing the model and ensures that underrepresented stocks retain relevance. In all our experiments, we set $w_k$ to be $1$. The integration of local models into a global model enables the detection of market-wide patterns that extend beyond individual stock movements. By aggregating trends from multiple stocks, the global model learns shared patterns such as correlated price fluctuations, responses to macroeconomic indicators, and sectoral dependencies. These cross-stock relationships improve predictive robustness, particularly in highly dynamic market conditions where isolated stock models may struggle to generalize. 

One of the challenges in global model merging is ensuring that the aggregation process does not overly smooth individual stock characteristics, which may lead to loss of stock-specific signals. To mitigate this, we employ an iterative refinement process where the merged global model is evaluated after each aggregation step to monitor performance convergence. The merging continues until the global model stabilizes, ensuring that it retains both cross-stock insights and essential stock-specific variations. In our experiments, we initially applied equal weights to all individual models, assuming uniform importance across stocks. This baseline approach demonstrated stable performance; however, further refinements can be introduced by dynamically adjusting $ w_k$ using market-specific criteria. Beyond improving model accuracy, the global model merging process enhances the stability of stock price predictions. Since individual stocks often experience periods of high volatility, sectoral shocks, or external economic impacts, a well-merged global model leverages the diversification effect to mitigate the impact of such anomalies. The iterative refinement ensures that the model does not become overly biased toward a subset of stocks while maintaining an accurate and generalized understanding of financial market behavior. By integrating insights across multiple stocks and refining the aggregation process through adaptive weighting, the global model effectively enhances predictive power compared to traditional single-stock forecasting. The resulting model not only improves overall accuracy but also adapts dynamically to evolving financial conditions, making it a robust framework for large-scale financial forecasting applications.

\subsection{Fine-Tuning for Stock-Specific Adaptation} 
Once the global model achieves sufficient convergence, the final phase involves fine-tuning the global model on individual stock data to ensure that stock-specific characteristics are preserved while leveraging the broader knowledge embedded in the global model. Although the global model provides a generalized understanding of market-wide trends, each stock exhibits unique behavioral patterns that may not be fully captured in a purely aggregated model. Fine-tuning addresses this limitation by reintroducing stock-specific learning while maintaining the benefits of cross-stock knowledge integration. The fine-tuning process begins by re-initializing the parameters of each local model $ \theta_k $ using the global model parameters $ \theta_g $. This re-initialization allows the local models to start from an optimized baseline rather than learning from scratch, significantly improving training efficiency and accelerating convergence. The stock-specific adaptation is then performed through further optimization on the individual stock dataset, refining the parameters to minimize the stock-level prediction loss:
\begin{align}
\theta_k^{\prime} = \arg\min_{\theta_k} \frac{1}{T} \sum_{t=1}^{T} \ell(f_k(\mathbf{X}_{k, t}; \theta_k), y_{k, t}).
\end{align}
where $ \ell $ represents the loss function, $ \mathbf{X}_{k, t} $ is the feature vector for stock $ S_k $ at time step $ t $, and $ y_{k, t} $ is the corresponding ground truth stock price. The fine-tuning process ensures that the model maintains the insights gained from the global model while adapting to the idiosyncratic features of each stock, preventing excessive bias toward overall market behavior.

A critical challenge in this phase is achieving the optimal balance between global generalization and stock-specific adaptation. If the fine-tuning step is too aggressive, the model may overfit to individual stock data, losing the generalization benefits gained from the global model. Conversely, if the fine-tuning is too minimal, the model may fail to capture the distinct price movement patterns of each stock. To address this, we introduce an adaptive learning rate strategy that controls the extent of fine-tuning. Specifically, we initialize the learning rate $ \eta_k $ for each stock $ S_k $ using a scaled version of the global learning rate $ \eta_g $:
\begin{align}
\eta_k = \alpha \cdot \eta_g,
\end{align}
where $ \alpha $ is a tuning coefficient that determines the adaptation strength. A lower $ \alpha $ ensures a smoother transition from the global model, preserving shared market trends, whereas a higher $ \alpha $ allows greater flexibility for local adaptation. Another important aspect of fine-tuning is the regularization of model parameters to prevent divergence from the global model while still allowing necessary stock-specific refinements. We incorporate an $ L_2 $ regularization term that penalizes excessive deviation from $ \theta_g $:
\begin{align}
\mathcal{L}_k^{\prime} = \frac{1}{T} \sum_{t=1}^{T} \ell(f_k(\mathbf{X}_{k, t}; \theta_k), y_{k, t}) + \lambda \|\theta_k - \theta_g\|^2,
\end{align}
where $ \lambda $ is a regularization coefficient that controls the trade-off between maintaining the global model structure and adapting to stock-specific features. This constraint ensures that while fine-tuning personalizes the model for individual stocks, the learned parameters remain aligned with the global market representation, thus avoiding drastic deviations that could lead to overfitting. Beyond parameter fine-tuning, feature-level adaptation can be incorporated to enhance stock-specific learning. Fine-tuning is crucial for ensuring that each stock model benefits from the cross-stock knowledge integrated in the global model while preserving the unique dynamics that define its price behavior. The resulting fine-tuned models exhibit improved accuracy in predicting stock movements compared to models trained in isolation, as they incorporate both market-wide trends and individual price variations. This final adaptation phase plays a pivotal role in making the \ac{CSTI} framework effective for real-world financial forecasting, balancing generalization and stock-specific refinement for optimal predictive performance.

\begin{table*}[ht!]
\setlength{\tabcolsep}{4pt}
\begin{center}
\caption{Experiment results using different training strategies over seven models on \ac{FNSPID} dataset with or without sentiment information. The results highlighted in red are the better ones. }
\label{tab:results}
\begin{tabular}{c | c | c c c c c c | c c c c c c }
\hline
\multirow{3}{*}{\#} & \multirow{3}{*}{Model} & \multicolumn{6}{c|}{Normal} & \multicolumn{6}{c}{Ours} \\
\cline{3-14}
& & \multicolumn{3}{c|}{w/ sentiment} & \multicolumn{3}{c|}{w/o sentiment} & \multicolumn{3}{c|}{w/ sentiment} & \multicolumn{3}{c}{w/o sentiment} \\
\cline{3-14}
& & MAE & MSE & \multicolumn{1}{c|}{$R^2$} & MAE & MSE & $R^2$ & MAE & MSE & \multicolumn{1}{c|}{$R^2$} & MAE & MSE & $R^2$ \\
\hline

\multirow{7}{*}{5} & Transformer &0.0702 & 0.0079 & \multicolumn{1}{c|}{ 0.3067 } & 0.0472 & 0.0045 & 0.5829  & \cellcolor{red!10}0.0535 & \cellcolor{red!10}0.005 & \multicolumn{1}{c|}{ \cellcolor{red!10}0.5243 } & \cellcolor{red!10} 0.0406 & \cellcolor{red!10}0.0031 & \cellcolor{red!10}0.6126 \\

& TimesNet &0.1158 & 0.0368 & \multicolumn{1}{c|}{ 0.444 } & 0.0459 & 0.0034 & 0.5368  &\cellcolor{red!10}0.0702 & \cellcolor{red!10}0.0095 & \multicolumn{1}{c|}{ \cellcolor{red!10}0.4669 } & \cellcolor{red!10}0.0427 & \cellcolor{red!10}0.0029 & \cellcolor{red!10} 0.6166\\

& DLinear & 0.0424 & 0.0029 & \multicolumn{1}{c|}{ 0.6014 } & 0.0359 & 0.0021 & 0.7211  &\cellcolor{red!10}0.0216 & \cellcolor{red!10}0.0008 & \multicolumn{1}{c|}{ \cellcolor{red!10}0.8772 } & \cellcolor{red!10}0.0278 & \cellcolor{red!10}0.0013 & \cellcolor{red!10}0.825 \\

& FreTS &0.0361 & 0.0022 & \multicolumn{1}{c|}{ 0.6864 } & 0.0308 & 0.0016 & 0.7579  &\cellcolor{red!10}0.0323 & \cellcolor{red!10}0.0017 & \multicolumn{1}{c|}{  \cellcolor{red!10}0.7499 } & \cellcolor{red!10}0.0311 & \cellcolor{red!10}0.0016 & \cellcolor{red!10}0.7625 \\

& PaiFilter & 0.0489 & 0.0037 & \multicolumn{1}{c|}{0.5051 } & 0.0489 & 0.0037 & 0.5062  &\cellcolor{red!10}0.0484 & 0\cellcolor{red!10}.0037 & \multicolumn{1}{c|}{ \cellcolor{red!10}0.511 } & \cellcolor{red!10}0.0388 & \cellcolor{red!10}0.0026 & \cellcolor{red!10}0.6026\\

& TexFilter & 0.0424 & 0.0029 & \multicolumn{1}{c|}{ 0.622 } & 0.0407 & 0.0026 & 0.653  &\cellcolor{red!10}0.0258 & \cellcolor{red!10}0.0014 & \multicolumn{1}{c|}{ \cellcolor{red!10}0.7588 } & \cellcolor{red!10}0.023 & \cellcolor{red!10}0.0009 & \cellcolor{red!10}0.8706 \\

& PatchTST & \cellcolor{red!10}0.0206 & \cellcolor{red!10}0.0008 & \multicolumn{1}{c|}{ \cellcolor{red!10}0.8467 } & 0.027 & 0.0012 & 0.8299  &0.0235 & 0.001 & \multicolumn{1}{c|}{ 0.8228 } & \cellcolor{red!10}0.0262 & \cellcolor{red!10}0.0012 & \cellcolor{red!10}0.8371 \\

\hline
\hline

\multirow{7}{*}{25} & Transformer & 0.0381 & 0.0036 & \multicolumn{1}{c|}{ 0.5946 } & \cellcolor{red!10}0.0489 & \cellcolor{red!10}0.0047 & \cellcolor{red!10}0.5298  & \cellcolor{red!10}0.0373 & \cellcolor{red!10}0.0031 & \multicolumn{1}{c|}{ \cellcolor{red!10}0.6585 } & 0.0583 & 0.0066 & 0.4212 \\

& TimesNet & \cellcolor{red!10}0.0341 & \cellcolor{red!10}0.0021 & \multicolumn{1}{c|}{ \cellcolor{red!10}0.6664 } & \cellcolor{red!10}0.0333 & \cellcolor{red!10}0.0021 & \cellcolor{red!10}0.7023  & 0.0358 & 0.0021 & \multicolumn{1}{c|}{ 0.6388 } & 0.0318 & 0.0017 & 0.7005 \\

& DLinear &  0.0229 & 0.001 & \multicolumn{1}{c|}{ 0.7776 } & 0.0259 & 0.0011 & 0.7605  & \cellcolor{red!10}0.0179 & \cellcolor{red!10}0.0007 & \multicolumn{1}{c|}{ \cellcolor{red!10}0.9225 } & \cellcolor{red!10}0.0223 & \cellcolor{red!10}0.0009 & \cellcolor{red!10}0.8344 \\

& FreTS &  \cellcolor{red!10}0.0195 & \cellcolor{red!10}0.0007 & \multicolumn{1}{c|}{ \cellcolor{red!10}0.868 } & 0.0255 & 0.0011 & 0.7858  & 0.0271 & 0.0013 & \multicolumn{1}{c|}{ 0.7793 } & \cellcolor{red!10}0.0253 & \cellcolor{red!10}0.0011 & \cellcolor{red!10}0.8029 \\

& PaiFilter &  0.0433 & 0.0031 & \multicolumn{1}{c|}{ 0.4405 } & 0.0433 & 0.0031 & 0.4398  & \cellcolor{red!10}0.0143 & \cellcolor{red!10}0.0005 & \multicolumn{1}{c|}{ \cellcolor{red!10}0.9125 } & \cellcolor{red!10}0.0165 & \cellcolor{red!10}0.0005 & \cellcolor{red!10}0.9096 \\

& TexFilter &  0.0218 & 0.0009 & \multicolumn{1}{c|}{ 0.7964 } & 0.0289 & 0.0014 & 0.7123  & \cellcolor{red!10}0.0122 & \cellcolor{red!10}0.0003 & \multicolumn{1}{c|}{ \cellcolor{red!10}0.9412 } & \cellcolor{red!10}0.0167 & \cellcolor{red!10}0.0005 & \cellcolor{red!10}0.8954 \\

& PatchTST &  \cellcolor{red!10}0.0143 & \cellcolor{red!10}0.0004 & \multicolumn{1}{c|}{ \cellcolor{red!10}0.9198 } & \cellcolor{red!10}0.023 & \cellcolor{red!10}0.0009 & \cellcolor{red!10}0.8382  & 0.0168 & 0.0005 & \multicolumn{1}{c|}{ 0.9045 } & 0.0254 & 0.0011 & 0.8007 \\

\hline
\hline

\multirow{7}{*}{50} & Transformer & \cellcolor{red!10}0.027 & \cellcolor{red!10}0.0017 & \multicolumn{1}{c|}{ \cellcolor{red!10}0.8162 } & \cellcolor{red!10}0.0536 & \cellcolor{red!10}0.0066 & \cellcolor{red!10}0.5834  & 0.0404 & 0.0045 & \multicolumn{1}{c|}{ 0.653 } & 0.0474 & 0.0051 & 0.5432\\

& TimesNet & \cellcolor{red!10}0.0283 & \cellcolor{red!10}0.0015 & \multicolumn{1}{c|}{ \cellcolor{red!10}0.7673 } & \cellcolor{red!10}0.0224 & \cellcolor{red!10}0.001 & \cellcolor{red!10}0.8449  &0.0303 & 0.0016 & \multicolumn{1}{c|}{ 0.7552 } & 0.037 & 0.0025 & 0.6613 \\

& DLinear & 0.0168 & 0.0005 & \multicolumn{1}{c|}{ 0.909 } & 0.0237 & 0.001 & 0.8368  & \cellcolor{red!10}0.0154 & \cellcolor{red!10}0.0005 & \multicolumn{1}{c|}{\cellcolor{red!10}0.936 } & \cellcolor{red!10}0.0234 & \cellcolor{red!10}0.001 & \cellcolor{red!10}0.8437 \\

& FreTS & \cellcolor{red!10}0.0266 & \cellcolor{red!10}0.0031 & \multicolumn{1}{c|}{ \cellcolor{red!10}0.8783 } & 0.0276 & 0.0014 & 0.7959  & 0.0249 & 0.0011 & \multicolumn{1}{c|}{ 0.8143 } & \cellcolor{red!10}0.0273 & \cellcolor{red!10}0.0013 & \cellcolor{red!10}0.8 \\

& PaiFilter & 0.041 & 0.0028 & \multicolumn{1}{c|}{ 0.5432 } & 0.0411 & 0.0028 & 0.5425  &\cellcolor{red!10}0.0375 & \cellcolor{red!10}0.0024 & \multicolumn{1}{c|}{ \cellcolor{red!10}0.6176 } & \cellcolor{red!10}0.0313 & \cellcolor{red!10}0.0018 & \cellcolor{red!10}0.7232 \\

& TexFilter & 0.0148 & 0.0005 & \multicolumn{1}{c|}{ 0.9223 } & 0.0233 & 0.001 & 0.8362  & \cellcolor{red!10}0.0123 & \cellcolor{red!10}0.0003 & \multicolumn{1}{c|}{\cellcolor{red!10}0.9498 } & \cellcolor{red!10}0.017 & \cellcolor{red!10}0.0005 & \cellcolor{red!10}0.9112 \\

& PatchTST & \cellcolor{red!10}0.0149 & \cellcolor{red!10}0.0004 & \multicolumn{1}{c|}{ \cellcolor{red!10}0.9325 } & \cellcolor{red!10}0.0247 & \cellcolor{red!10}0.0011 & \cellcolor{red!10}0.8351  & 0.0164 & 0.0005 & \multicolumn{1}{c|}{ 0.9186 } & 0.0254 & 0.0011 & 0.8219 \\

\hline
\end{tabular}
\end{center}
\end{table*}

The \ac{CSTI} framework presents multiple advantages over traditional single-stock training methods by combining local learning with global trend integration. The parallel training of individual models significantly enhances computational efficiency, allowing the system to scale across a vast number of stocks. The global model merging phase enables learning from inter-stock dependencies, mitigating the limitations of isolated training and improving generalization across different market conditions. By merging individual models into a global representation, \ac{CSTI} captures shared financial trends, such as sector-based correlations and macroeconomic influences, leading to enhanced prediction accuracy. Additionally, the fine-tuning phase ensures that while benefiting from global insights, each stock model retains unique price movement patterns, improving stock-specific forecasts. Unlike conventional \ac{FL} approaches, which focus solely on privacy and decentralization, the \ac{CSTI} framework is explicitly designed to improve the predictive power of financial models by leveraging cross-stock information. This method ensures that stock-specific characteristics are preserved, while broader financial trends are effectively captured, making \ac{CSTI} an ideal solution for large-scale financial applications.


\section{Experiments}
\label{sec:er}

\subsection{Settings and Dataset}

We utilized the PyTorch framework~\cite{paszke2019pytorch} to implement all neural network models. For reproducibility, the source code is publicly available and can be accessed at the following 
link\footnote{\url{https://github.com/Rand2AI/CSTI}}
. Local training was conducted using various optimization algorithms, all of which employed a consistent learning rate of $0.01$ and a momentum of $0.9$. The local batch size was set to $64$, and the number of training epochs was fixed at $100$. For our proposed strategy, we trained local models for $1$ epoch before integrating them to generate a global model during the first $50$ epochs. Following this, we fine-tuned the global model for the remaining $50$ epochs. This approach ensures that both the normal training process and the \ac{CSTI} framework have the same total number of model weight update iterations. To evaluate the performance of our method across different stock group sizes, we randomly selected three groups of subdatasets, containing $5$, $25$, and $50$ stocks, respectively. These experiments were conducted to assess the scalability and performance of our method with varying numbers of stocks. The backbone models used in our experiments include Transformer, TimesNet, DLinear, FreTS, PaiFilter, TexFilter, and PatchTST.

In our study, we conducted experiments using \ac{FNSPID}, which is a comprehensive dataset designed to address key limitations in financial market analysis by integrating numerical stock data with sentiment-laden financial news. This dataset spans the years $1999$ to $2023$, encompassing over $29.7$ million stock price entries and $15.7$ million time-aligned financial news records for $4775$ companies within the $S\&P500$ index. The dataset uniquely combines quantitative and qualitative data to enhance predictive modeling capabilities, particularly in the realm of stock price forecasting. \ac{FNSPID} was built through a rigorous process combining data scraping, cleaning, and integration. Stock price data was sourced from Yahoo Finance APIs, while financial news was collected via web scraping from platforms like NASDAQ. Ethical considerations, including adherence to copyright laws and avoidance of premium content, were prioritized throughout data collection. Sentiment scores were normalized and missing data gaps were handled through exponential decay methods to maintain temporal consistency. It sets a new standard for financial datasets, offering a rich resource for advancing research in financial market analysis, sentiment modeling, and stock price prediction. Its scale, diversity, and integration of sentiment data make it an invaluable tool for both academic and practical applications. In out experiments, we used two sets of data features. The first set includes Open and Close prices and the second one includes Open, Close and Sentiment. 

\subsection{Results Analysis}

\begin{figure*}[ht!]
\centering
    \subfigure[Normal, Transformer]{
    \centering
    \includegraphics[width=0.32\linewidth]{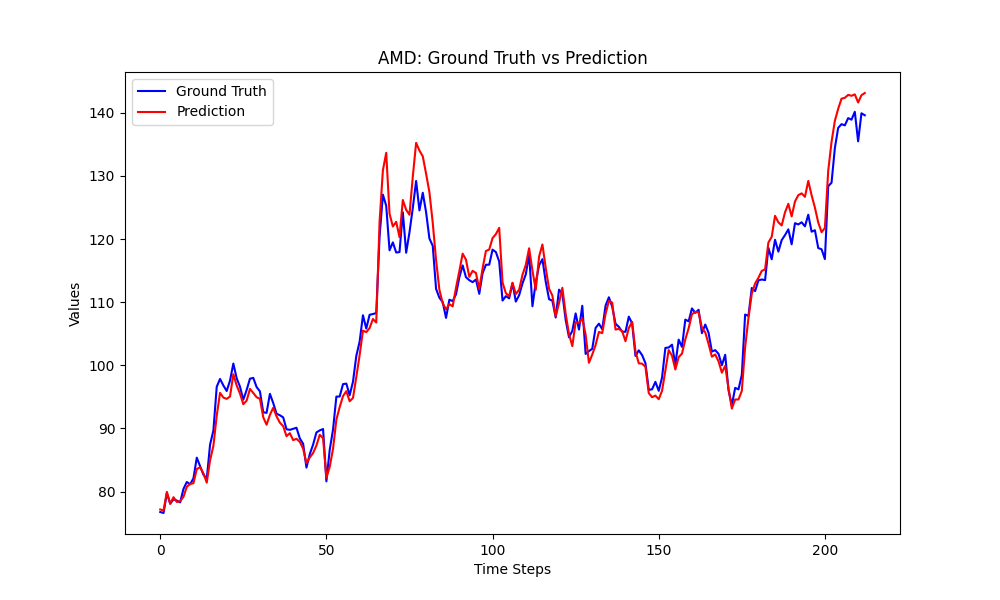}
    }
    \hspace{-.35in}
    \subfigure[CSTI, Transformer]{
    \centering
    \includegraphics[width=0.32\linewidth]{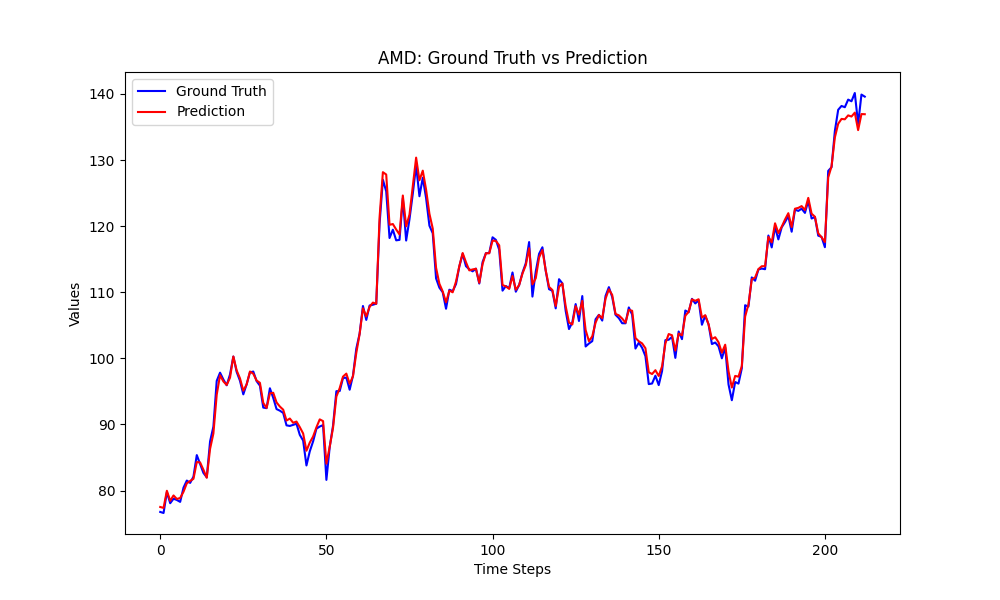}
    }
    \hspace{-.35in}
    \subfigure[CSTI pre-trained, Transformer]{
    \centering
    \includegraphics[width=0.32\linewidth]{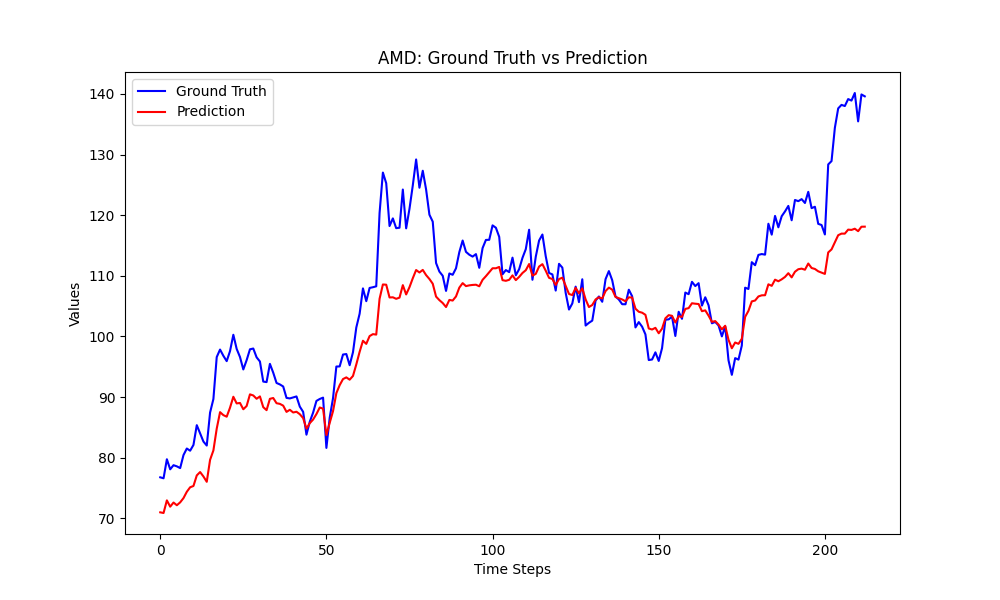}
    }
    
    \subfigure[Normal, TimesNet]{
    \centering
    \includegraphics[width=0.32\linewidth]{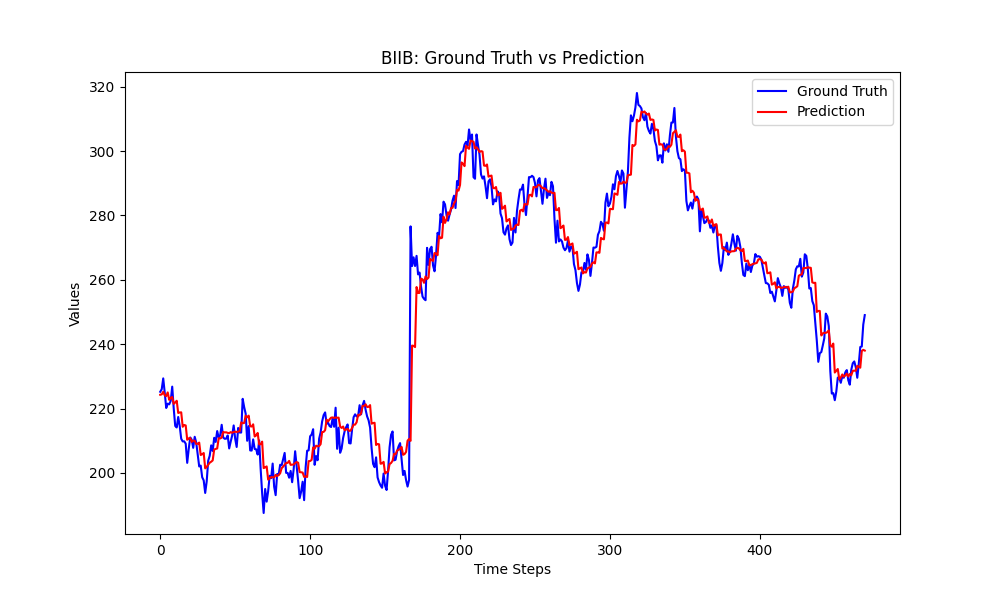}
    }
    \hspace{-.35in}
    \subfigure[CSTI, TimesNet]{
    \centering
    \includegraphics[width=0.32\linewidth]{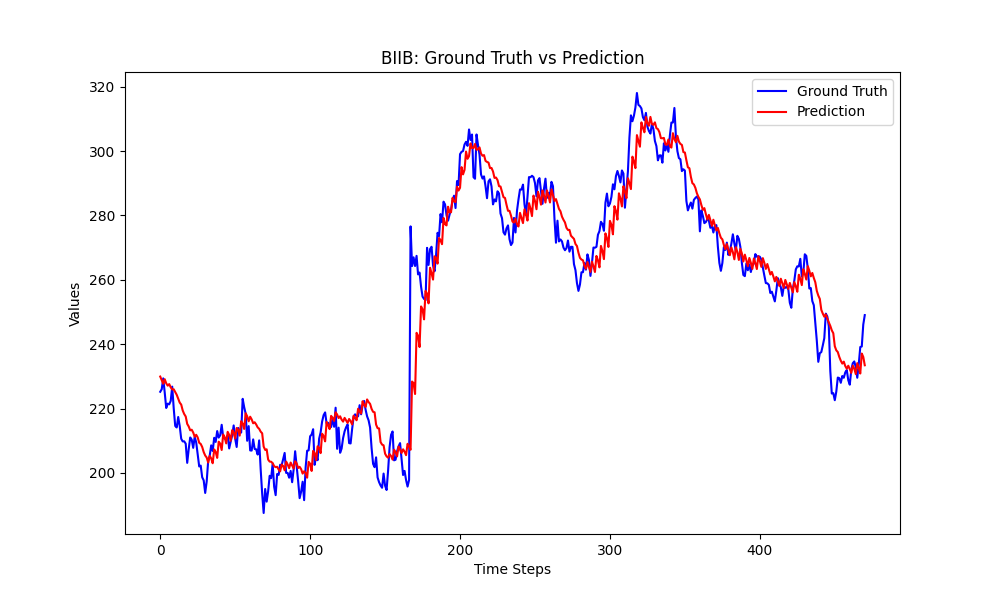}
    }
    \hspace{-.35in}
    \subfigure[CSTI pre-trained, TimesNet]{
    \centering
    \includegraphics[width=0.32\linewidth]{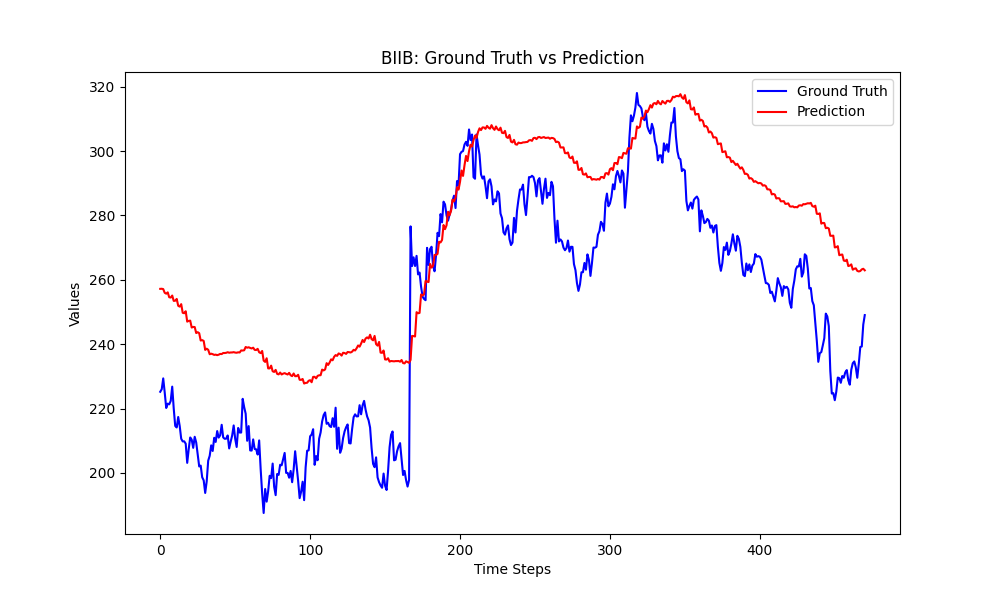}
    }
    
    \subfigure[Normal, DLinear]{
    \centering
    \includegraphics[width=0.32\linewidth]{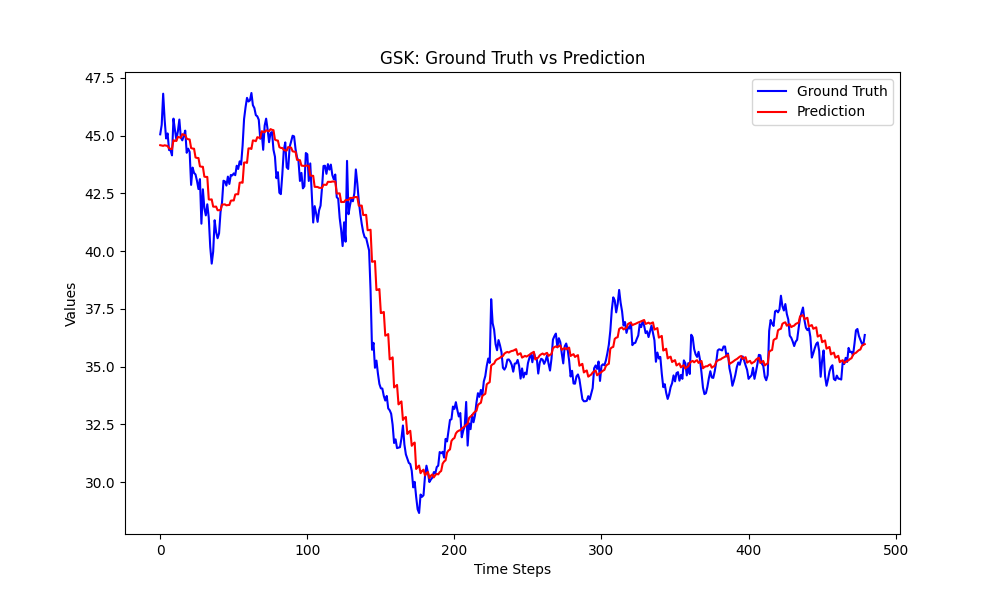}
    }
    \hspace{-.35in}
    \subfigure[CSTI, DLinear]{
    \centering
    \includegraphics[width=0.32\linewidth]{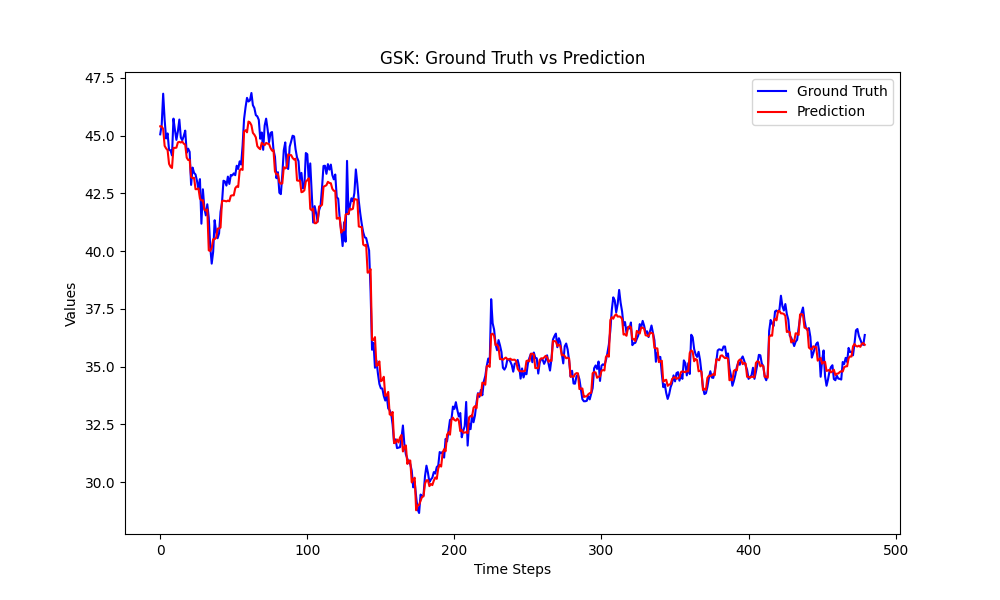}
    }
    \hspace{-.35in}
    \subfigure[CSTI pre-trained, DLinear]{
    \centering
    \includegraphics[width=0.32\linewidth]{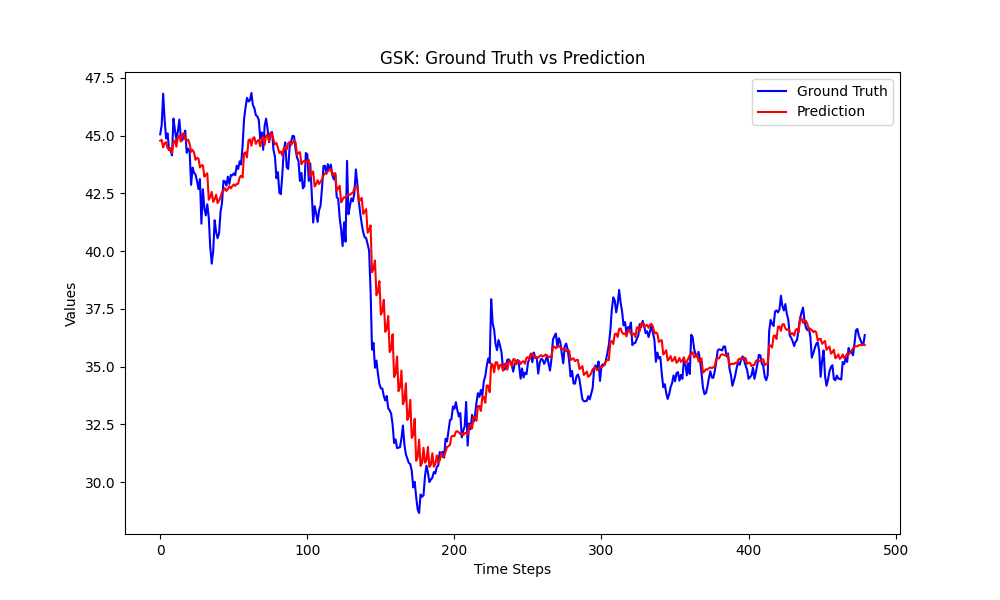}
    }

    \subfigure[Normal, PatchTST]{
    \centering
    \includegraphics[width=0.32\linewidth]{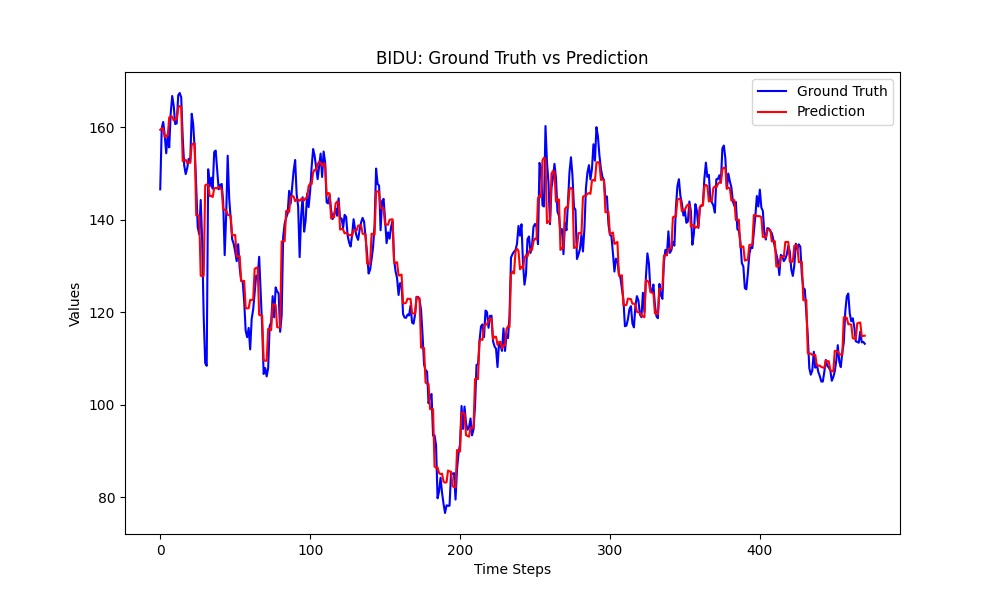}
    }
    \hspace{-.35in}
    \subfigure[CSTI, PatchTST]{
    \centering
    \includegraphics[width=0.32\linewidth]{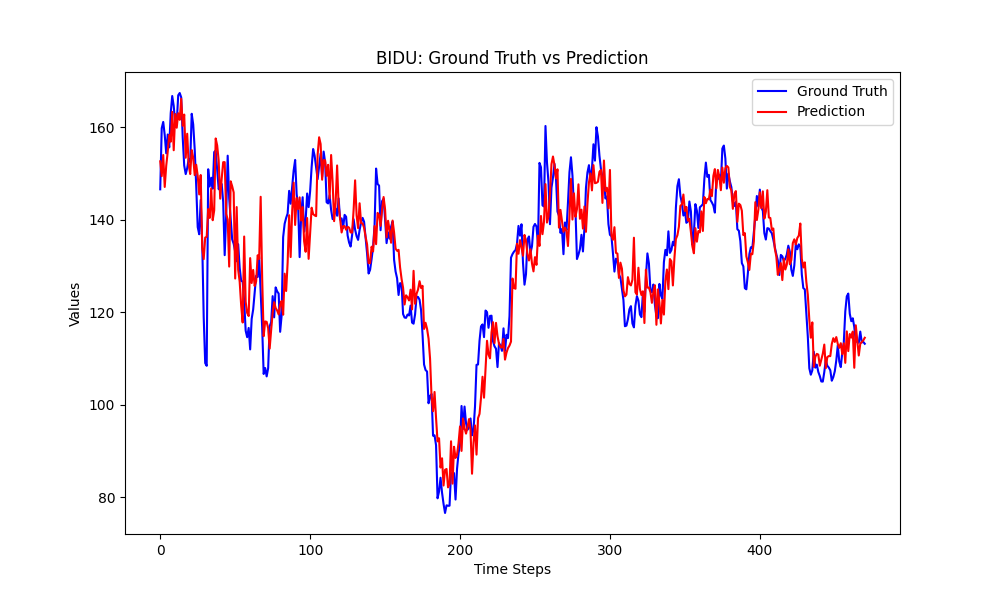}
    }
    \hspace{-.35in}
    \subfigure[CSTI pre-trained, PatchTST]{
    \centering
    \includegraphics[width=0.32\linewidth]{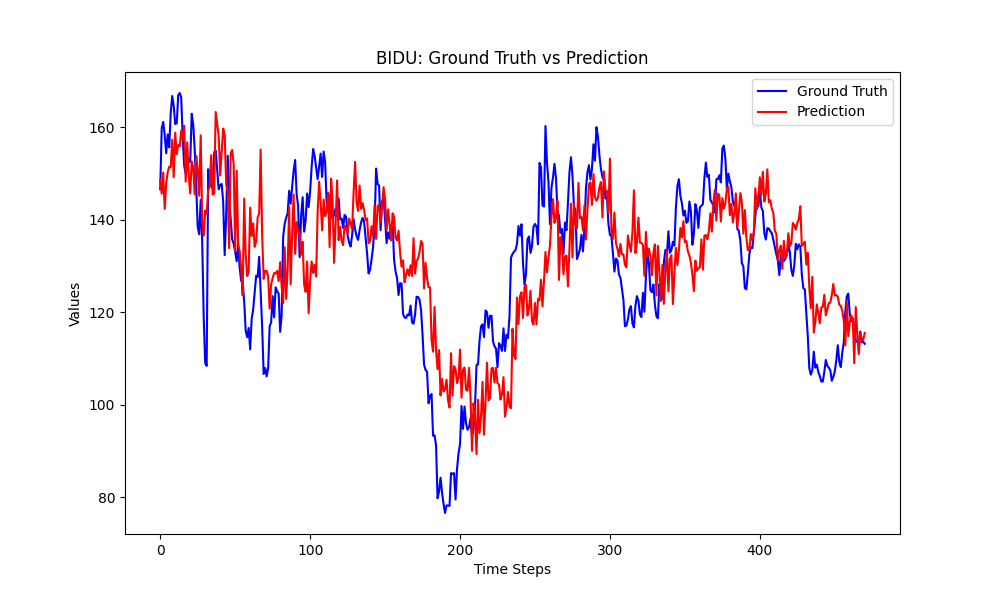}
    }

    \subfigure[Normal, TexFilter]{
    \centering
    \includegraphics[width=0.32\linewidth]{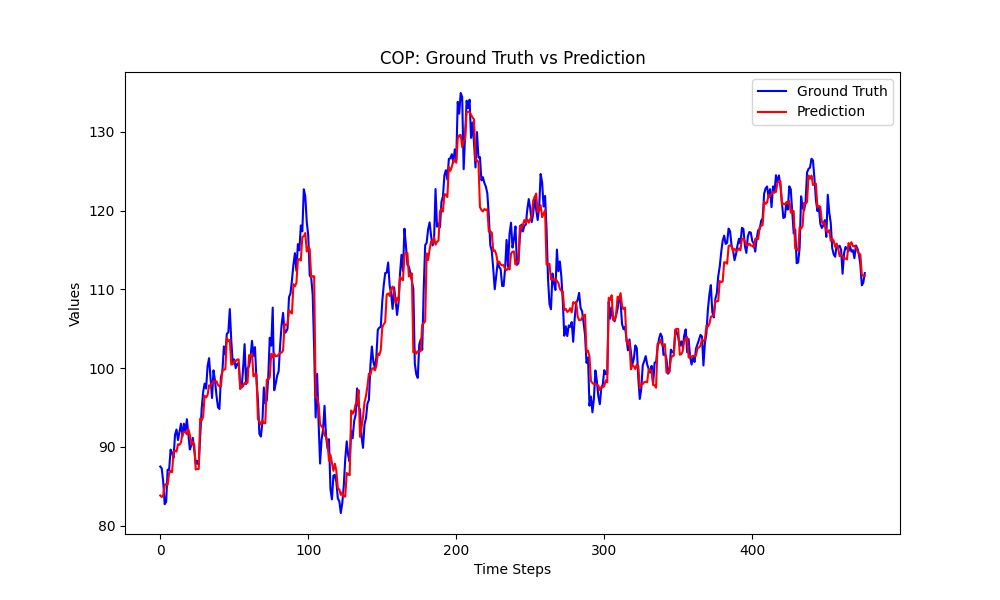}
    }
    \hspace{-.35in}
    \subfigure[CSTI, TexFilter]{
    \centering
    \includegraphics[width=0.32\linewidth]{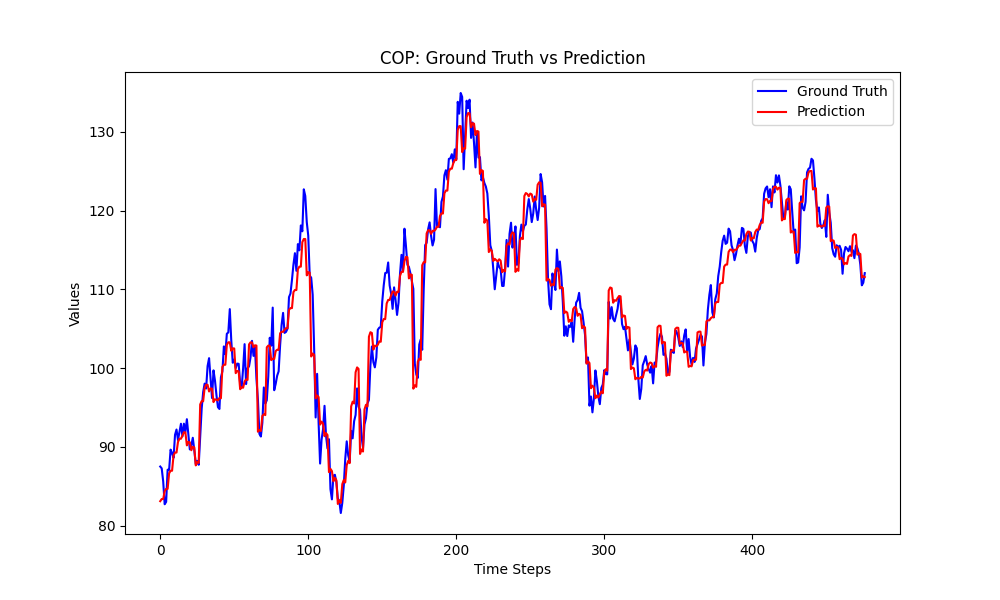}
    }
    \hspace{-.35in}
    \subfigure[CSTI pre-trained, TexFilter]{
    \centering
    \includegraphics[width=0.32\linewidth]{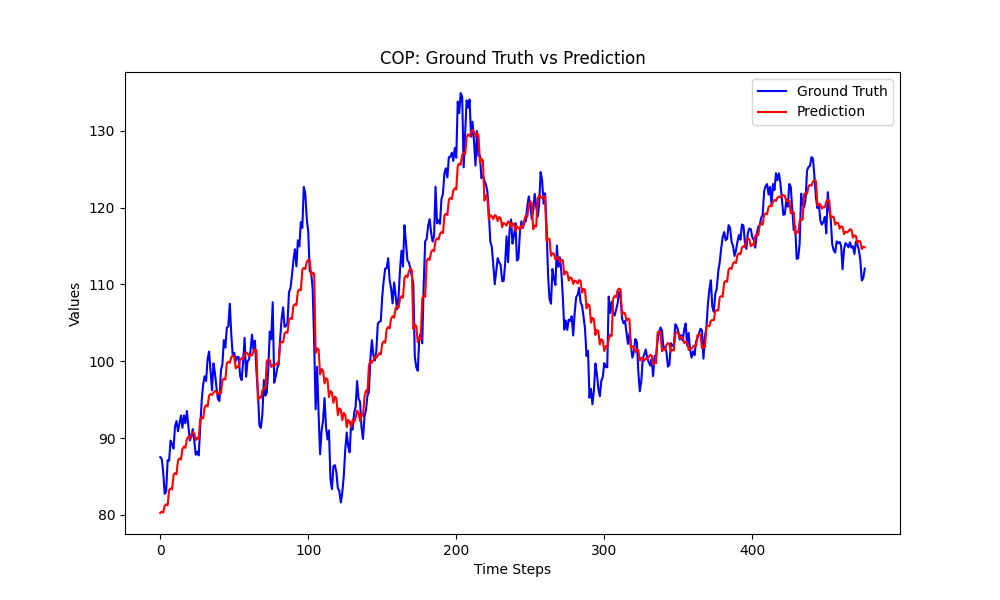}
    }
\caption{The visualization of the predicted regression lines for models trained on 50 stocks without sentiment information is presented. The first column shows the results from the normal training strategy, the second column presents the results from our proposed methods, and the third column displays the results from pre-trained global models.}
\label{fig:rl}
\vspace{.10in}
\end{figure*}

\begin{figure*}[ht!]
    \centering
    \includegraphics[width=0.99\linewidth]{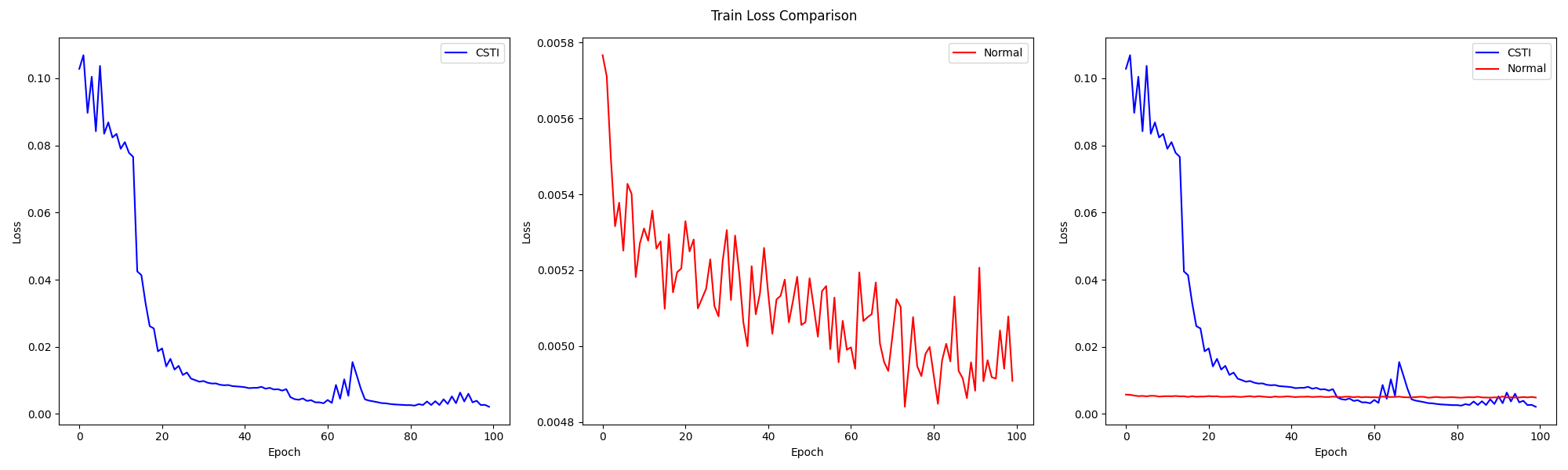}
    \caption{Training loss comparison using Transformer pre-trained on 50 stocks data and fine-tuned on AMD stock prices.}
    \label{fig:loss}
\end{figure*}

A set of comprehensive experiments were conducted to evaluate the performance of our proposed method compared to the normal training strategy. Following the work in \cite{dong2024fnspid}, the normal training strategy involves training a model on one stock's historical price data, then fine-tuning it on the next stock's data, iteratively, until all the stock data has been utilized. Table~\ref{tab:results} presents the experimental results for seven models on the \ac{FNSPID} dataset, with and without sentiment information. The results highlighted in red indicate the superior ones. It is evident that our proposed training strategy, \ac{CSTI}, outperforms the normal training strategy in most experiments. Notably, for the PaiFilter model, our method achieved $R^2$ values of $0.9125$ and $0.9096$ on data with and without sentiment information, respectively, using $25$ stocks. In contrast, the normal training strategy only achieved $R^2$ values of $0.4405$ and $0.4398$ under the same conditions. However, our method does not always yield better results. For instance, in the case of TimesNet and PatchTST, the normal training strategy outperformed our approach across all experiments using $25$ and $50$ stocks. Additionally, models trained on data with sentiment information consistently outperformed those trained on data without sentiment information, supporting the claim made in~\cite{dong2024fnspid}.

In Figure~\ref{fig:rl}, we present visualizations of some predicted regression lines from the normal training strategy, our proposed \ac{CSTI} training strategy, and \ac{CSTI} pre-trained global models. The models were trained on $50$ stocks without sentiment information. The first column displays results from the normal training strategy, the second column shows results from our proposed \ac{CSTI} methods, and the third column presents results from \ac{CSTI} pre-trained global models. Figures (a), (b), and (c) depict the performance of the Transformer model on AMD stock. While the normal training strategy (Figure (a)) and the \ac{CSTI} method (Figure (b)) closely follow the ground truth, the pre-trained model (Figure (c)) diverges slightly in local patterns but effectively captures the overall increasing and decreasing trends of the stock price. Figures (d), (e), and (f) illustrate the results for the TimesNet model on BIIB stock. The normal training strategy (Figure (d)) achieves a better fit to fluctuations compared to the \ac{CSTI} method (Figure (e)), while the pre-trained global model (Figure (f)) shows the ability to track major trends learned during the cooperative learning stage but lacks the precision achieved by fine-tuning. Figures (g), (h), and (i) compare the DLinear model's predictions on GSK stock. The \ac{CSTI} method (Figure (h)) demonstrates reduced noise and improved alignment with the ground truth compared to the normal training strategy (Figure (g)), while the pre-trained global model (Figure (i)) captures general trends but deviates from the exact latent patterns. Figure (j), (k), and (l) give the performance of the PatchTST model. In this case, the normal training (Figure (j)) also performs better than \ac{CSTI}. In the last row, Figure (m), (n) and (o), present the performance from TexFilter model. It is hard to distinguish which method does a better job. Comparing the results from the \ac{CSTI} method and its pre-trained models, we observe that the global model effectively learned the overall increasing and decreasing trends of stock prices during the cooperative learning stage. In the fine-tuning stage, the model further refined its understanding of the latent stock price trends, producing accurate and precise prediction results. These comparisons emphasize the effectiveness of the proposed \ac{CSTI} training strategy in improving prediction accuracy and generalization across diverse stock trends while highlighting the complementary benefits of pre-training and fine-tuning in leveraging global knowledge and domain-specific patterns.

Figure~\ref{fig:loss} illustrates the training loss over 100 epochs for two Transformer models: one trained using the normal training strategy (red line) and the other trained with our proposed \ac{CSTI} strategy (blue line). Both experiments were conducted using $50$ stocks without sentiment information as the dataset. The normal training strategy exhibits a steady and smooth decline in training loss from the very first epoch. This is expected, as the Transformer model quickly adapts to the stock price data, which is relatively straightforward to regress. As a result, the initial loss is already quite low, and it decreases further in a gradual and consistent manner. In contrast, the \ac{CSTI} approach shows a training loss pattern more typical of general \ac{DL} models, with a sharper decline in the initial epochs followed by minor oscillations as training progresses. This behavior is due to the integration of $50$ local models into a single global model at each epoch, which introduces variability and prevents the initial loss from being as small as that observed in normal training. Despite this, the \ac{CSTI} strategy achieves a similar low level of training loss as the normal training method by the end of the $100$ epochs, demonstrating its effectiveness in converging to an optimal solution. The figure highlights the differences in loss dynamics between the two strategies, with \ac{CSTI} achieving comparable final performance while leveraging cross-stock integration to capture broader market relationships.
\section{Conclusion}
\label{sec:cc}

In conclusion, this study highlights the transformative potential of integrating cross-stock patterns into stock price prediction models using the proposed \ac{CSTI} framework. Traditional methods, while effective, often neglect the inter-dependencies among stocks, leading to suboptimal utilization of available market information. By leveraging a novel training strategy that iteratively merges local stock models into a unified global model, \ac{CSTI} demonstrates its ability to enhance prediction accuracy and generalization. Extensive experiments conducted on the \ac{FNSPID} dataset validate the effectiveness of the method, showcasing superior performance across various models and settings. The proposed approach not only addresses the limitations of single-stock training but also integrates complementary benefits of pre-training and fine-tuning to capture both global trends and stock-specific nuances. The results underscore \ac{CSTI}’s scalability, efficiency, and robustness, paving the way for innovative advancements in financial forecasting.

\bibliographystyle{IEEEtran}
\bibliography{ref}

\newpage

\begin{IEEEbiography}[{\includegraphics[width=1in,height=1.25in,clip,keepaspectratio]{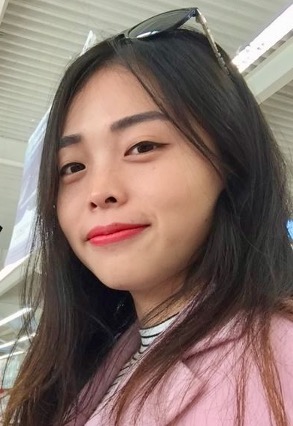}}]{Yi Hu}
received the M.Sc. in Mathematical Computing from Swansea University. She is currently a computer science Ph.D. student at Swansea University. Her research area is federated learning.
\end{IEEEbiography}
\begin{IEEEbiography}[{\includegraphics[width=1in,height=1.25in,clip,keepaspectratio]{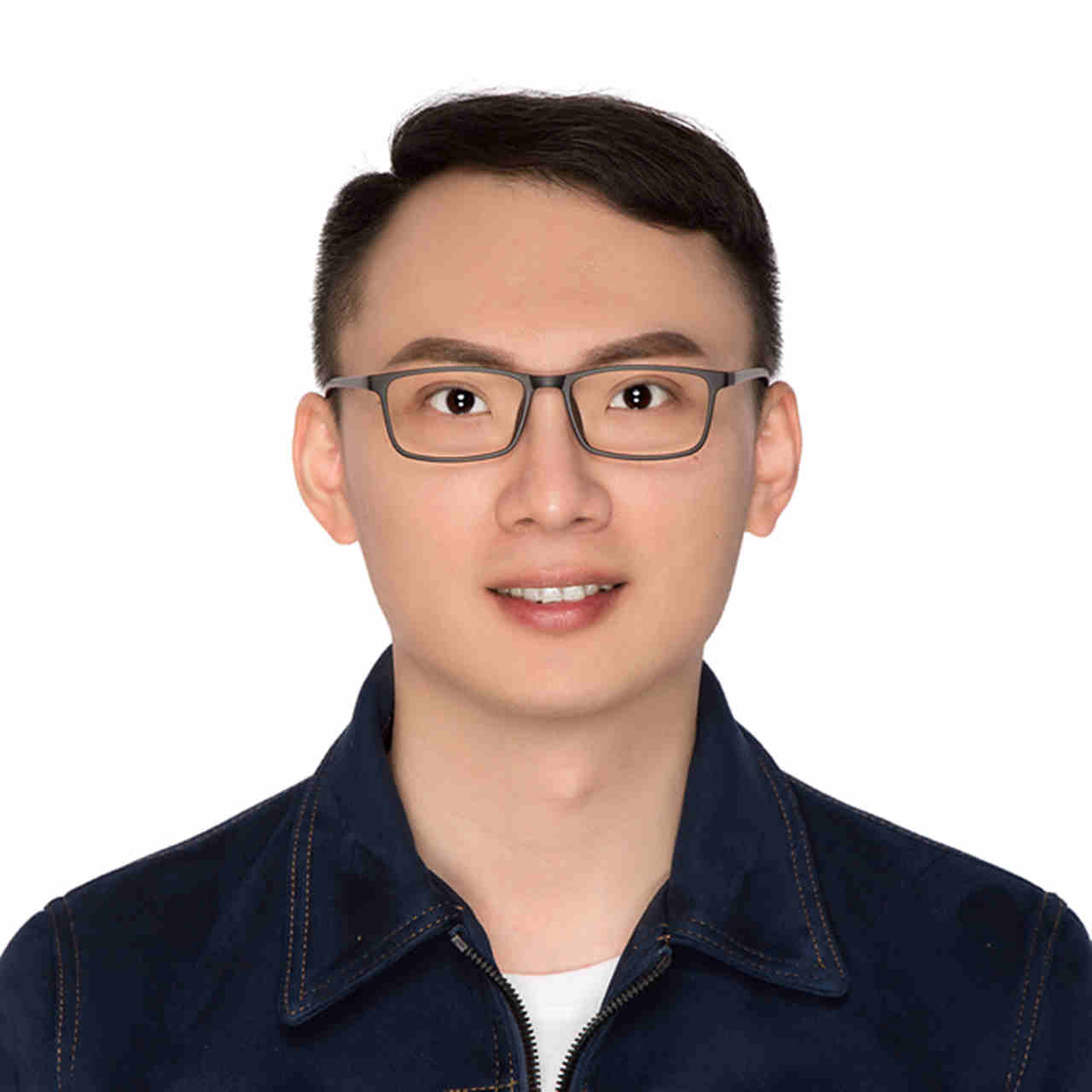}}]{Hanchi Ren}
received the M.Sc. and Ph.D. degree in computer science from Swansea University, U.K., in 2016 and 2023. He is currently an academic tutor in the Computer Vision and Machine Learning Laboratory, Department of Computer Science, Swansea University. His research subject is on privacy-preserving federated learning and machine learning and artificial intelligence in general and applications in computer vision and robotics.
\end{IEEEbiography}
\begin{IEEEbiography}[{\includegraphics[width=1in,height=1.25in,clip,keepaspectratio]{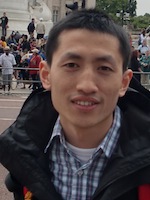}}]{Jingjing Deng}
received his Ph.D. in Visual Computing from Swansea University, UK, in 2017. Presently, he holds the position of Assistant Professor in the Department of Computer Science at Durham University, UK. He founded the Rand2AI Lab in 2022 which actively engages in cutting-edge research in computer vision and artificial intelligence. In recent years, the team has focused on developing computational models that can cultivate and generalize intelligence from and for the complex world.
\end{IEEEbiography}
\begin{IEEEbiography}[{\includegraphics[width=1in,height=1.25in,clip,keepaspectratio]{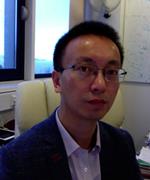}}]{Xianghua Xie}
received the M.Sc. and Ph.D. degrees in computer science from the University of Bristol, Bristol, U.K., in 2002 and 2006, respectively. He is currently a Full Professor with the Department of Computer Science, Swansea University, Swansea, U.K., and is leading the Computer Vision and Machine Leaning Laboratory, Swansea University. He has published more than 160 refereed conference and journal publications and (co-)edited several conference proceedings. His research interests include various aspects of pattern recognition and machine intelligence and their applications to real-world problems. He is a member of BMVA. He is an Associate Editor of a number of journals, including Pattern Recognition and IET Computer Vision.
\end{IEEEbiography}

\vfill

\end{document}